# DRU-NET: AN EFFICIENT DEEP CONVOLUTIONAL NEURAL NETWORK FOR MEDICAL IMAGE SEGMENTATION

*Mina Jafari[1], Dorothee Auer[2], Susan Francis[3], Jonathan Garibaldi[1], Xin Chen[1]*

[1]Intelligent Modeling and Analysis Group, School of Computer Science, University of Nottingham, UK
[2]School of Medicine, University of Nottingham, UK
[3]Sir Peter Mansfield Imaging Centre, University of Nottingham, UK

## ABSTRACT

Residual network (ResNet) and densely connected network (DenseNet) have significantly improved the training efficiency and performance of deep convolutional neural networks (DCNNs) mainly for object classification tasks. In this paper, we propose an efficient network architecture by considering advantages of both networks. The proposed method is integrated into an encoder-decoder DCNN model for medical image segmentation. Our method adds additional skip connections compared to ResNet but uses significantly fewer model parameters than DenseNet. We evaluate the proposed method on a public dataset (ISIC 2018 grand-challenge) for skin lesion segmentation and a local brain MRI dataset. In comparison with ResNet-based, DenseNet-based and attention network (AttnNet) based methods within the same encoder-decoder network structure, our method achieves significantly higher segmentation accuracy with fewer number of model parameters than DenseNet and AttnNet. The code is available on GitHub (GitHub link: https://github.com/MinaJf/DRU-net).

## 1. INTRODUCTION

Earlier semantic image segmentation approaches, before the arrival of deep learning, rely on hand-crafted features and their combination with classifiers for pixel-level classification. However, the performance of these systems has always been limited by the robustness of the designed feature descriptors. Since 2012, based on the idea of deep convolutional neural network (DCNN) proposed by LeCun [1], the performance of computer vision systems have been significantly improved in a wide range of applications. In this paper, we focus on DCNN based solutions for the problem of medical image segmentation.

Nowadays, encoder-decoder [2], fully convolutional network (FCN) [3] and dilated convolutional networks [4] are state-of-the-art image segmentation models. U-net [2], which is trained in an encoder-decoder architecture, has outperformed previous works in terms of the number of required training samples, memory and computational time. Several variants of U-net have been proposed, which mainly focus on improving segmentation accuracy and efficiency of feature information passing within and across layers. For instance, H-DenseU-net [5] combines the idea of densely connected network [6] and U-net for 3D liver and tumour segmentation. Alternatively, U-net++ [7] connects encoder to decoder by using dense skip connections between different layers. MDU-net [8] is proposed to add three multi-scale dense connections in U-net simultaneously, i.e. dense encoder, decoder and the connection between them. FU-net [9] modifies U-net by proposing a dynamically weighted cross-entropy loss function.

Generally, based on previous studies, the encoder-decoder DCNN architecture has shown a superior performance against other architectures due to its capability of capturing features in a multi-scale manner. Most methods achieve better performance by integrating sophisticated network blocks (e.g. dense network, attention network [10], etc.), which require more parameters to be learned. However, it would be more desirable to reduce the number of model parameters and still achieve a similar or better performance. Therefore, in this paper, we compare the performance of residual network (ResNet) [11] and dense network (DenseNet) [6], and further propose some modifications to improve the efficiency of the network. As the main contribution of this paper, we propose a simple but efficient network block which is able to achieve better image segmentation performance than DenseNet, ResNet and an attention net based method [10]. More importantly, our method requires fewer model parameters than these methods (except for ResNet).

## 2. METHODOLOGY

We choose U-net [2] as the basic structure for performance comparison of different network blocks for image segmentation. Additional to the original U-net, a batch normalization (BN) operation [12] is added to each layer, as it is a well-known strategy to achieve faster convergence and enable stabilized network training.

Different from conventional Conv-ReLU operation, ResNet [11] adds a shortcut connection from the input to the output of a Conv-ReLU-Conv-ReLU process as illustrated in Fig. 1-(a). Furthermore, DenseNet architecture [6], which has outperformed most previous networks (including ResNet) in ILSVRC challenge, consists of several interconnected dense-blocks. Each dense-block has a classical Conv-BN-ReLU process and a "bottle neck" block, as illustrated in Fig. 1-(b). The input of each dense-block is a concatenated feature map of the outputs of all previous dense-blocks together with the original input of the first dense-block, as illustrated by the top diagram in Fig. 1-(b).

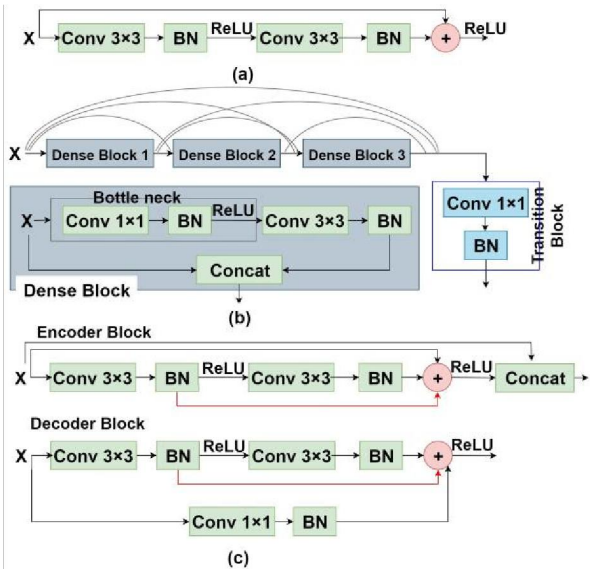

**Fig. 1:** a) Residual block. b) Dense net blocks. c) Encoder and decoder blocks in DRU-net.

One drawback of ResNet, compared with DenseNet, is the lack of dense concantenation of all previous convolutional outputs to the successive feature maps. The original ResNet contains two Conv operations inside the skip connection (Fig. 1-(a)). If gradients vanish at the second Conv during backpropagation, the parameters in the first Conv can not be updated neither. In contrast, DenseNet utilizes information from all previous convolutional operations, which allows gradients to flow through several paths and enables richer information to be combined for feature extraction in concecutive layers. However, as shown in Fig. 1-(b), it requires extra "bottle neck" and "transition" blocks to reduce the feature map channels. This effectively performs feature map aggregation in a learnable way. However, these additional feature aggregation steps lead to extra parameters and heavier computational load, especially when several layers of DenseNet are used. By considering the advantages and disadvantages of these two networks, we propose to add an additional connection between the output of first Conv-BN operations to the last Conv-BN output (red shortcut connection in Fig. 1-(c)) with a summation operation for feature map aggregation, as illustrated in Fig. 1-(c). The justification behind the idea is that the additional shortcut connection allows the parameters to be updated in the first Conv even the gradients in the second Conv approach zeros. This is sufficient to allow the gradients to be backpropagated efficiently. Unlike the DenseNet that uses multiple learnable 1×1 Conv to aggregate feature maps, we use a simple summation to combine feature maps (same as ResNet). It can be observed from the evaluation results (Section 3.2) that using these learnable aggregation operations in DenseNet cannot lead to a better segmentation performance but with more parameters and longer training time.

Additionally, for each layer in the encoder path of our method, a concatenation is applied to combine the input and output (see Encoder Block in Fig. 1(c)). The combined feature map is then feed into the next layer. This concatenation is not only to include more information from the input of each layer but also to make the dimension of feature maps directly compatible with the feature maps in the next layer without extra parameters (unlike the transition block in DenseNet). Moreover, in the decoder block as shown in Fig. 1-(c), a Conv1×1 is used to reduce the number of channels of input instead of cropping it to make the dimension compatiable to the layer output for a summation operation. In a general case, if more than two Conv-BN blocks are used, the outputs from all Conv-BN operations need to be added to the final Conv-BN output. This would not increase the number of parameters and computational time. By integrating the proposed encoder and decoder blocks into the U-net architecture, we name our proposed segmentation method as DRU-net (Dense Residual U-net). Based on the above modifications, the overall scheme of DRU-net is illustrated in Fig. 2.

## 3. EXPERIMENTS AND RESULTS

In this section, material, experimental design and network parameters are described. Based on ISIC 2018 challenge dataset for lesion segmentation [13], the proposed DRU-net is compared with the original U-net [2], ResNet-based U-net (RU-net), DenseNet-based U-net (DU-net) and an attention network based method [10] (AttnU-net) in terms of segmentation accuracy measured by Dice coefficient (DC), Jaccard, precision and recall. We also use a challenging brain MRI dataset to demonstrate the performance of our method by varying the size of training data in a multi-class segmentation task.

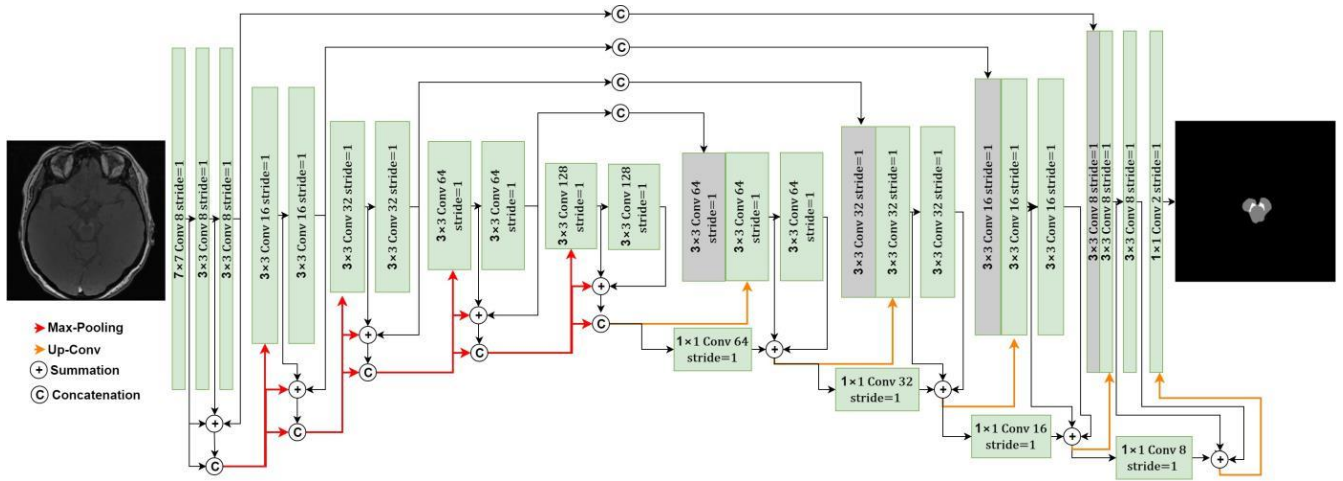

**Fig. 2:** The overall scheme of DRU-net.

### 3.1. Material and Network Parameter Settings

#### 3.1.1. ISIC 2018

Skin Lesion Analysis Toward Melanoma Detection grand challenge dataset [13] consists of 2594 RGB images of skin lesions with binary annotations. For our experiments, the images were firstly converted to gray level and resized to 192×256 pixels with a split of 75%-25% for training and testing (same as in [10]).

#### 3.1.2. Brain Data

T1-weighted brain MRIs for segmentation of Midbrain (MB) and Substantia Nigra (SN) [14] were acquired in Nottingham Hospital and were approved by the local ethics committee for this research. The dataset is composed of 102 subjects each has 30 axial image slices. 3 or 4 slices were manually selected by a radiologist that contain both the MB and SN. We have a total number of 310 2D slices for our experiments. An example pair of original brain image (input) and the segmented image (output) is shown in Fig. 2. In the segmented image, three classes are presented: background (black), MB (gray), and SN (white). This dataset is challenging due to the extremely unbalanced number of pixels in each class. We performed three experiments for each method using different training/ testing sizes. 10 images were randomly extracted first for model validation.

The remaining 300 images were randomly split into 200 training/ 100 testing, 100 training/ 200 testing and 50 training/ 250 testing. This allows the comparison of different methods when the size of training data varies.

#### 3.1.3. Network Settings

The number of layers was 5 for each of the encoder and decoder paths for all compared methods. For U-net, RU-net, DU-net and our method, cross entropy loss function was used and Adam optimizer [15] was applied with an initial learning rate of 0.001. As described in the original paper [10], AttnU-net used a new loss function based on Tversky index with three hyper parameters. It also added attention blocks and optimized by stochastic gradient descent with momentum. For the ISIC 2018 dataset, all methods were trained for 50 epochs with a batch size of 4. All methods for brain data were trained with batch size of 2 for 200 epochs.

### 3.2. RESULTS

Table 1 lists the results for the skin lesion dataset. The results show that U-net produced the lowest DC and Jaccard values. RU-net achieved better performance than U-net for DC, Jaccard and precision measures. AttnU-net and DU-net produced similar results in terms of DC and Jaccard measures, which were better than U-net and RU-net. DRU-net significantly (statistically significant using Wilcoxon signed rank test with p<0.001) outperformed all other methods for DC, Precision and Jaccard measures. AttnU-net achieved the highest recall. It may be due to that the settings of the three hyper-parameters for the Tversky loss function favour the recall measurement (see [10] for details).

Some visual results of using U-net, RU-net, DU-net and DRU-net are presented in Fig. 3. It is seen that the segmentation result of DRU-net (Fig. 3-(f)) is more similar to the ground truth mask (Fig. 3-(b)) than the other methods, especially for the region around the lesion boundary. The segmentation result of U-net (Fig. 3-(c)) and RU-net (Fig. 3-(d)) fail to recover precise lesion boundaries. DU-net (Fig. 3-(e)) had a more precise segmentation result than U-net and RU-net, but still produced more false positives than DRU-net around the lesion boundary.

**Table 1.** Comparison of U-net, RU-net, DU-net, AttnU-net and DRU-net on ISIC 2018. The mean dice coefficient values, precision, recall, and Jaccard index are reported.

| Method | DC | Precision | Recall | Jaccard |
|---|---|---|---|---|
| U-net [2] | 0.840 | 0.865 | 0.869 | 0.724 |
| RU-net [11] | 0.848 | 0.892 | 0.857 | 0.743 |
| DU-net [6] | 0.855 | 0.894 | 0.865 | 0.748 |
| AttnU-net [10] | 0.856 | 0.858 | **0.897** | 0.748 |
| **DRU-net (ours)** | **0.861** | **0.919** | 0.882 | **0.755** |

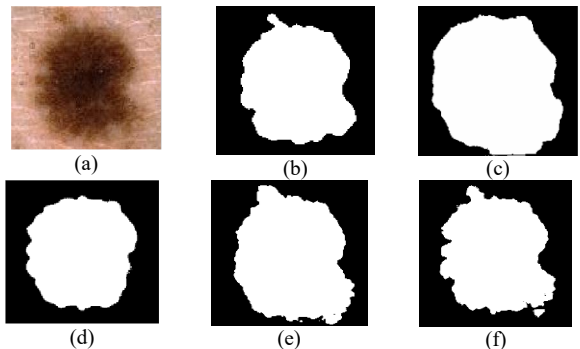

**Fig. 3**: a) Original image. b) Ground truth. c) Segmentation result by U-net. d) Segmentation result by RU-net. e) Segmentation result by DU-net. f) Segmentation result by DRU-net.

We also compared the methods in a multi-class segmentation task using the brain MRI data for MB and SN segmentation. Segmentation accuracy (i.e. DC) and training time are reported in Table 2 for different experiments by varying the size of training set as described in Section 3.1.2. We conclude from Table 2 that DRU-net was significantly better than RU-net and U-net for both MB and SN classes despite of training size. DRU-net was significantly better than DU-net when the number of training images were reduced to 50 for MB class. The segmentation of SN class is challenging as there are very few pixels in this class. Our proposed DRU-net outperformed all other methods with statistical significance in using 100 and 50 training samples for the SN segmentation. The performance differences became more significant when the training size was decreased. In terms of training time, U-net, RU-net and DRU-net were similar and significantly faster than DU-net. This is due to the larger number of parameters required by the DU-net for the "bottle neck" and "transition" blocks. The number of parameters for RU-net, DU-net and DRU-net were ~2 million, ~6 million and ~3 million respectively.

Therefore, it is obvious that the proposed DRU-net is an efficient network in terms of memory, training time and segmentation accuracy.

**Table 2.** Comparison of U-net, RU-net, DU-net and DRU-net on different sizes of training data using brain MRI dataset. The mean dice coefficient values with Wilcoxon signed rank test results and training time are reported. Based on Wilcoxon signed rank test: * indicates result of DRU-net is different from U-net with statistical significance (p<0.01); ¬ indicates result of DRU-net is different from RU-net with statistical significance (p<0.01); ^ indicates result of DRU-net is different from DU-net with statistical significance (p<0.01).

| Number of training / testing samples | Method | DC | | Training Time (Min.) |
|---|---|---|---|---|
| | | MB | SN | |
| 200/100 | U-net | 0.8850 | 0.7383 | 117.69 |
| | RU-net | 0.9068 | 0.7782 | 148.87 |
| | DU-net | **0.9167** | 0.7908 | 238.36 |
| | **DRU-net** | 0.9132*¬ | **0.7956*¬** | 164.63 |
| 100/200 | U-net | 0.8473 | 0.7018 | 68.53 |
| | RU-net | 0.8578 | 0.7126 | 70.24 |
| | DU-net | **0.8764** | 0.7572 | 136.62 |
| | **DRU-net** | 0.8643*¬^ | **0.7713*¬^** | 96.62 |
| 50/250 | U-net | 0.7984 | 0.5877 | 68.53 |
| | RU-net | 0.8101 | 0.6439 | 58.90 |
| | DU-net | 0.8220 | 0.6578 | 90.78 |
| | **DRU-net** | **0.8428*¬^** | **0.7240*¬^** | 50.64 |

## 4. DISCUSSION AND CONCLUSIONS

We have presented a simple but efficient network that is designed to take the advantages of DenseNet and ResNet. Additional minor modifications like aggregation of feature maps using summation and concatenation of the input to the output at each layer make the feature learning process more efficient. By evaluating the methods on a public skin lesion dataset and a local brain MRI dataset, our proposed DRU-net has shown to outperform U-net, RU-net and DU-net significantly, especially for the label class that has small number of pixels and with small number of training examples. For the skin lesion dataset, our method also outperformed one of the state-of-the-art methods using attention network and Tversky loss function. We only demonstrated the efficiency of our proposed network in the context of image segmentation within an encoder-decoder DCNN structure. The applicability and efficiency of applying the proposed network to other scenarios (e.g. object classification) with larger datasets are worth exploring in future work.